\documentclass{article}
\usepackage{spconf,amsmath,graphicx}
\usepackage[skip=5pt]{caption}
\usepackage{cite}
\usepackage{acronym}
\usepackage{amssymb,amsfonts}
\usepackage{amsmath}
\DeclareMathOperator*{\argmax}{arg\,max}

\usepackage{algorithmic}
\usepackage{graphicx}
\usepackage{xcolor}
\usepackage{url}
\usepackage{balance}
\usepackage{tabularx}
\usepackage{booktabs,multirow} 

\usepackage{placeins}
\usepackage{pifont}
\usepackage{float}
\usepackage{multirow}
\usepackage{caption}
\usepackage{subcaption}
\usepackage[T1]{fontenc}
\usepackage[colorlinks=true,citecolor=blue,allcolors=blue]{hyperref}

\acrodef{HRTF}{Head-Related Transfer Function}
\acrodef{HRIR}{Head-Related Impulse Response}
\acrodef{TrF}{Transfer Function}
\acrodef{TSE}{Target Speaker Extraction}
\acrodef{TF}{Time-Frequency}
\acrodef{GMM}{Gaussian Mixture Model}
\acrodef{CGMM}{Complex Gaussian Mixture Model}
\acrodef{WDO}{W-Disjoint Orthogonality}
\acrodef{MLE}{Maximum Likelihood Estimation}
\acrodef{MSE}{Mean Square Error}
\acrodef{RMSE}{Root Mean Square Error}
\acrodef{EM}{Expectation-Maximization}
\acrodef{REM}{Recursive EM}
\acrodef{CREM}{Capp{\'e} and Moulines REM}
\acrodef{RIR}{Room Impulse Responses}
\acrodef{STFT}{Short-Time Fourier transform}
\acrodef{ReLU}{Rectified Linear Unit}
\acrodef{DOA}{Direction of Arrival}
\acrodef{TDOA}{Time Difference of Arrival}
\acrodef{SI-SDR}{Scale Invariant Signal to Distortion Ratio }
\acrodef{FFT}{Fast Fourier Transform}
\acrodef{BSS}{Blind Source Separation}
\acrodef{SSL}{Sound Source Localization}
\acrodef{GCC}{Generalized Cross Correlation }
\acrodef{MUSIC}{Multiple Signals Classification}
\acrodef{SRP-PHAT}{Steered Response Power with Phase Transform}
\acrodef{MAE}{Mean Absolute Error}
\acrodef{WSJ}{Wall Street Journal}
\acrodef{SOFA}{Spatially Oriented Format for Acoustics}
\acrodef{SR}{Sampling Rate}
\acrodef{SNR}{Signal to Noise Ratio}
\acrodef{IR}{Impulse Response}
\acrodef{PESQ}{Perceptual Evaluation of Speech Quality}
\acrodef{MOS}{Mean Opinion Score}
\acrodef{ILD}{Interaural Level Difference}
\acrodef{ITD}{Interaural Time Difference}
\acrodef{STOI}{Short-time Objective Intelligibility}
\acrodef{ATF}{Acoustic Transfer Function}
\acrodef{RTF}{Relative Transfer Function}
\acrodef{RT}{Reverberant Time}
\acrodef{PRP}{Pair-Wise Relative Phase Ratio}
\acrodef{DNN}{Deep Neural Network}
\acrodef{DL}{Deep Learning}
\acrodef{dB}{Decibel}
\acrodef{NN}{Neural Network}
\acrodef{FC}{Fully Connected}
\acrodef{SIR}{Signal-to-Interference Ratio}
\acrodef{p.d.f.}{Probability Density Function}
\acrodef{LCMV}{Linearly Constrained Minimum Variance}
\acrodef{BLCMV}{Binaural Linearly Constrained Minimum Variance}
\acrodef{LS}{LibriSpeech}
\acrodef{MLS}{Multilingual LibriSpeech}
\acrodef{MIMO}{Multi input Multi output}
\acrodef{Bi-TSE}{Binaural Target Speaker Extraction}
\acrodef{Bi-TSE-HRTF}{Binaural Target Speaker Extraction using HRTFs}
\acrodef{RI}{Real-Imaginary}
\acrodef{MOS}{Mean Opinion Score}
\acrodef{FiLM}{Feature-wise Linear Modulation}
\acrodef{PE}{Positional Encoding}
\acrodef{MLP}{Multi-Layer Perceptron}

\title{Unfolded Recursive Expectation-Maximization Neural Network for Speaker Tracking}
\name{Rina Veler and Sharon Gannot}
\address{Faculty of Engineering, Bar-Ilan University, 
Ramat-Gan, Israel \\
\{velerri,sharon.gannot\}@biu.ac.il}
\begin{document}
\ninept
\maketitle

\begin{abstract}
We propose a deep unfolded \ac{REM} network for robust tracking of a single moving speaker in mild reverberant environments. Unlike classical \ac{REM} algorithms, which rely on fixed-step-size decay schedules, the proposed architecture learns an adaptive update policy by unfolding the iterative procedure into differentiable layers. We introduce a Step Size Network that leverages \ac{FiLM} and \ac{PE} to dynamically adjust the recursion weights based on temporal context and convergence state.
Experimental results for tracking a single speaker under reverberant conditions demonstrate that the proposed unfolded network outperforms the classical \ac{CREM} baseline, which employs a spatial grid search to map the estimated centroids to physical positions. In the single-speaker tracking task, the proposed method achieves a lower \ac{RMSE} than the \ac{CREM} baseline, highlighting its potential for dynamic acoustic scenarios.
\end{abstract}

\begin{keywords}
Sound source tracking; Unfolding neural network; Recursive Expectation-Maximization; Pair-wise relative phase ratio.
\end{keywords}
\section{Introduction}
\ac{SSL} is a fundamental task in modern audio applications, ranging from autonomous robots to smart-home consumer electronics. In realistic environments, estimating the spatial coordinates of a source is significantly complicated by reverberation and ambient noise. Traditional methods, such as \ac{SRP-PHAT}, \ac{MUSIC}, and \ac{MLE} \cite{Knapp1976GeneralizedCorrelation, Schmidt1986Multiple}, often struggle in these adverse conditions. This has motivated a shift toward \ac{DNN}-based approaches, which offer superior robustness in complex acoustic scenes \cite{Chakrabarty2019MultiSpeaker,Grumiaux2022Survey}.

However, purely data-driven \acp{DNN} are often criticized for their lack of interpretability. Algorithm \emph{unfolding} (or unrolling) bridges this gap by casting the iterations of classical algorithms as differentiable network layers \cite{Gregor2010Learning, Monga2021Algorithm}. In our previous work \cite{RinaVeler2026UnfoldedEM}, we successfully employed an unfolded network to estimate static multi-speaker positions by integrating the batch \ac{EM} procedure—previously used for \ac{PRP} clustering \cite{Schwartz2014, Mandel2007EMAlgorithm}—into an unrolled architecture. While effective for static sources, batch-oriented methods lack the temporal consistency required for dynamic speaker tracking.

To address moving-source scenarios, recursive variants of the \ac{EM} algorithm, e.g., the algorithm proposed by Cappé and Moulines \cite{Cappe2009OnlineEM} (henceforth denoted the \ac{CREM} algorithm), update parameters frame-by-frame using recursively accumulated sufficient statistics. Several studies have integrated \ac{EM} iterations into unfolding frameworks for image segmentation and clustering \cite{Greff2017NeuralEM, Pu2023DeepEM}. 
Although recursive, a major limitation in the application of \ac{CREM} is the selection of the step-size parameter $\gamma_t$, which governs the trade-off between past information and current observations. In static scenarios, a fixed decay schedule should be used. However, in tracking scenarios, where the speaker position changes over time, the step size should adapt to the underlying dynamics and is therefore often chosen empirically.

In this paper, we introduce a deep Unfolded \ac{CREM} network embedded within an encoder-decoder architecture that learns an adaptive, data-driven step-size policy. By incorporating a conditioning sub-network based on \ac{FiLM} \cite{perez2018film} and sinusoidal \ac{PE} \cite{vaswani2017attention}, our model dynamically adjusts the recursion weights based on the current tracking state and temporal context. We demonstrate that this hybrid approach improves tracking accuracy and convergence speed compared to the classical recursive method.

\section{Problem Formulation}
\label{Problem Formulation }

Consider a single speaker moving in a reverberant enclosure, whose activity is captured by an array of $M$ microphone pairs. We assume the speaker moves at a constant velocity along either a linear or a circular path. The analysis is carried out in the \ac{STFT} domain. The signal captured at the $j$th microphone of the pair $m$, where $j = 1,2$ and $m = 1,\ldots,M$, at time-frame $t$ and frequency bin $k$ is modeled as:
\begin{equation}
z_{m,j}(t,k) = a_{m,j}(t,k) \cdot s(t,k) + n_{m,j}(t,k)
\end{equation} 
where $s(t,k)$ is the source signal, $n_{m,j}(t,k)$ is additive noise, and $a_{m,j}(t,k)$ are the \acp{ATF} relating the source positions and the microphone positions. The \acp{ATF} vary over time as the speaker moves along its trajectory $\mathbf{p}(t)$.
Similar to our previous work \cite{Schwartz2014,RinaVeler2026UnfoldedEM}, we employ the \ac{PRP} as the localization feature to improve robustness to signal-power variations. The observed \ac{PRP} for pair $m$ is defined as:
\begin{equation}
\phi_m(t,k) \triangleq \frac{z_{m,2}(t,k)}{z_{m,1}(t,k)} \cdot \frac{|z_{m,1}(t,k)|}{|z_{m,2}(t,k)|},
\end{equation}
and the observation vector at each time-frequency bin is
\begin{equation}
\boldsymbol{\phi}(t,k) = \left[\phi_1(t,k), \dots, \phi_M(t,k)\right]^\top.
\end{equation}
To model the received PRP, we adopt a \ac{CGMM} with two components ($C=2$): one associated with the moving speaker ($c=1$) and the other with noise or spurious estimates ($c=2$). The \ac{p.d.f.} for a single observation $\boldsymbol{\phi}(t,k)$ is given by:
\begin{equation}
f(\boldsymbol{\phi}(t,k)) = \sum_{c=1}^{2} \psi_c(t) \cdot \mathcal{N}^c\left(\boldsymbol{\phi}(t,k); \tilde{\boldsymbol{\phi}}^k(\mathbf{p}_c(t)), \mathbf{\Sigma}_c\right)
\end{equation}
where $\psi_c(t)$ is the prior probability of the $c$-th cluster. For the speaker cluster ($c=1$), the mean $\tilde{\boldsymbol{\phi}}^k(\mathbf{p}(t))$ should be the expected \ac{PRP} analytically derived from the current position $\mathbf{p}(t)$, where each element $m$ is calculated as:
\begin{equation}\label{eq:PRP}
\tilde{\phi}_m^k(\mathbf{p}(t)) \triangleq \exp\left(-j\frac{2\pi k}{K} \frac{|\mathbf{p}(t) - \mathbf{p}_{m,2}| - |\mathbf{p}(t) - \mathbf{p}_{m,1}|}{v \cdot T_s}\right)
\end{equation}
where $v$ represents the sound velocity, and $T_s$ the sampling time of the continuous-time signal.
For the noise cluster ($c=2$), the mean $\tilde{\boldsymbol{\phi}}^k(\mathbf{p}(t))$ does not originate from a specific point source, and hence, we expect this cluster to have a larger variance compared to the speaker's variance.

Following \cite{Schwartz2014}, we assume independence of the \ac{PRP} features across microphone pairs.
We further simplify each covariance matrix to a scaled identity matrix, i.e., $\mathbf{\Sigma}_c = \sigma_c^2 \mathbf{I}_M$.
%
%
Finally, the \ac{p.d.f.} of the entire observation set is given by
\begin{multline} \label{eq:cgmm_likelihood}
f(\boldsymbol{\phi}; \boldsymbol{\theta}(t)) = \\\prod_{t,k} \left[ \sum_{c=1}^{2} \psi_c(t) \prod_{m=1}^M \mathcal{N}^c\left(\phi_m(t,k); \tilde{\phi}_m^k(\mathbf{p}_c(t)), \sigma_c^2(t)\right) \right],
\end{multline}
where $\boldsymbol{\theta}(t)$ denotes the set of time-varying parameters, comprising the Gaussian means, the priors, and variances.  
In the following section, we derive the \ac{REM} algorithm, which updates $\hat{\boldsymbol{\theta}}(t)$ by accumulating the score of the log-likelihood over time, and demonstrate how this procedure is unfolded into a deep neural network to learn optimal, time-adaptive update steps.

\section{Proposed Method}
\label{Proposed Method}
For online tracking of a moving speaker, we develop a recursive framework based on the \ac{CREM} algorithm, which we subsequently unfold into a deep neural network.

\vspace{3pt}\noindent\textbf{The EM Algorithm:}
\label{The EM Algorithm}
The \ac{EM} algorithm requires defining three components and their probability models: the observations ($\boldsymbol{\phi}$), the target parameters ($\boldsymbol{\theta}$), and the hidden data ($\mathbf{x}$). In our case, we define the hidden data $\mathbf{x}$ to be the association of each \ac{TF} bin with either the source at a particular location or the background noise.
Given the hidden data, the probability log-likelihood function of the observations is given by:
\sloppy
\begin{multline}
    f(\mathbf{x}, \boldsymbol{\phi}; \boldsymbol{\theta})  = \\\prod_{t,k}  \sum_{c} \psi_c\cdot x(t,k,c) \prod_{m} \mathcal{N}^c \left(\phi_m(t,k); \tilde{\phi}_m^k(\mathbf{p_c}), \sigma_c^2 \right).
\end{multline}
The E-step is therefore given by:
\begin{equation}
    Q(\boldsymbol{\theta}|\boldsymbol{\theta}^{(\ell-1)}) \triangleq E\left\{ \log f(\mathbf{x}, \boldsymbol{\phi}; \boldsymbol{\theta}) \Big| \boldsymbol{\phi}; \boldsymbol{\theta}^{(\ell-1)} \right\}.
\label{eq:e_step_def} 
\end{equation}
For implementing the E-step, it is sufficient to evaluate $\mu^{(\ell-1)}(t,k,c) \triangleq E\{x(t,k,c) | \boldsymbol{\phi}(t,k); \boldsymbol{\theta}^{(\ell-1)}\}$, given by:
\begin{equation}
\frac{\psi_c^{(\ell-1)} \prod_{m} \mathcal{N}^c \left(\phi_m(t,k); \tilde{\phi}_m^k(\mathbf{p}_c)^{(\ell-1)}, (\sigma_c^2)^{(\ell-1)} \right)}{\sum_{c} \psi_c^{(\ell-1)} \prod_{m} \mathcal{N}^c \left(\phi_m(t,k); \tilde{\phi}_m^k(\mathbf{p}_c)^{(\ell-1)}, (\sigma_c^2)^{(\ell-1)} \right)}.
\label{eq:posterior_prob} 
\end{equation}
Maximizing \eqref{eq:posterior_prob} with respect to the parameters $\boldsymbol{\theta}^{(\ell)}$ constitutes the M-step. We stress that this batch formulation assumes static speakers and operates entirely offline, making it unsuitable for real-time tracking in dynamic, time-varying environments.

\vspace{3pt}\noindent\textbf{The CREM Algorithm:}
\label{The CREM Algorithm}
The \ac{CREM} algorithm adapts the batch \ac{EM} procedure to an online scenario by substituting the iteration index with the time-frame index $t$. The core idea is to recursively estimate the auxiliary function, denoted $Q_R(\boldsymbol{\theta}|\boldsymbol{\theta}^{(t)})$, using a stochastic approximation step:
\begin{equation}
Q_R (\boldsymbol{\theta}|\boldsymbol{\theta}^{(t)}) = (1-\gamma_t)Q_R (\boldsymbol{\theta}|\boldsymbol{\theta}^{(t-1)}) + \gamma_t Q(\boldsymbol{\theta}|\boldsymbol{\theta}^{(t)}),
\label{eq:crem_q_recursion}
\end{equation}
where $\gamma_t$ is a time-varying smoothing coefficient (step-size) that determines the balance between past information and the current observation.
The complete data log-likelihood is given by:
\begin{multline}
\log f(\mathbf{x}_t, \boldsymbol{\phi}_t; \boldsymbol{\theta})=\sum_{k,c} x(t,k,c) \cdot \\\left(\log \psi_c(t) + \sum_{m} \log \mathcal{N}^c \left(\phi_m(t,k); \tilde{\phi}_m^k(\mathbf{p}_c(t)), \sigma_c^2(t) \right) \right).
\label{eq:log_likelihood_crem}
\end{multline}
Given that our \ac{CGMM} likelihood \eqref{eq:log_likelihood_crem} belongs to the exponential family, the maximization of $Q_R$ can be simplified by maintaining a recursive estimate of the sufficient statistics vector, $\boldsymbol{\eta}^R(t)$.\footnote{Due to space constraints, the derivation is significantly shortened.} The update rule for the sufficient statistics at frame $t$ is:
\begin{equation}
\boldsymbol{\eta}^R (t) = \boldsymbol{\eta}^R (t-1) + \gamma_{t} \left[ \bar{\boldsymbol{\eta}}(t) - \boldsymbol{\eta}^R (t-1) \right],
\label{eq:eta_recursion}
\end{equation}
where $\bar{\boldsymbol{\eta}}(t)$ is the expected sufficient statistics computed in the E-step, based on the posterior probability $\mu(t,k,c)$ that the $k$-th frequency bin at time $t$ belongs to cluster $c$. 
The term $\langle \boldsymbol{\overline{\eta}} (\mathbf{x}(t+1), \boldsymbol{\phi}(t+1)), \boldsymbol{\epsilon}(\boldsymbol{\theta}) \rangle$, required for the E-step,\footnote{ $\boldsymbol{\epsilon}(\boldsymbol{\theta})$ is related to the definition of the exponential family.} can be simplified by calculating:
\begin{equation}
   \mu(t+1,k,c)= E\{x(t+1,k,c)|\boldsymbol{\phi}_m(t+1,k);\boldsymbol{\theta}_{R}(t)\} 
\end{equation}
which is readily given by \eqref{eq:posterior_prob}. Using this definition, the expected sufficient statistic $\bar{\boldsymbol{\eta}}(\boldsymbol{\mu}(t), \boldsymbol{\phi}(t))$ can be written as:
\begin{equation}
\bar{\boldsymbol{\eta}}(\boldsymbol{\mu}(t), \boldsymbol{\phi}(t)) = \begin{pmatrix} \boldsymbol{\mu}(t) \\ \boldsymbol{\mu}(t) \odot |\boldsymbol{\phi}(t)|^2 \\ \boldsymbol{\mu}(t) \odot \boldsymbol{\phi}^*(t) \\ \boldsymbol{\mu}(t) \odot \boldsymbol{\phi}(t) \end{pmatrix}
\label{eq:eta_bar_vector} 
\end{equation}
where the Hadamard product $\odot$ is applied with implicit broadcasting; specifically, when a dimension is absent in one of the tensors, it is replicated along that dimension to match the size of the other, followed by element-wise multiplication.
The recursive sufficient statistic $\boldsymbol{\eta}^R (\boldsymbol{\mu}(t+1), \boldsymbol{\phi}(t+1))$ is defined as:
\begin{equation}
\boldsymbol{\eta}^R (\boldsymbol{\mu}(t+1), \boldsymbol{\phi}(t+1)) \triangleq \begin{pmatrix} \boldsymbol{\eta}_1^R (t+1) \\ \boldsymbol{\eta}_2^R (t+1) \\ \boldsymbol{\eta}_3^R (t+1) \\ \boldsymbol{\eta}_4^R (t+1) \end{pmatrix}
\label{eq:eta_r_vector} 
\end{equation}
where the updates are:
\begin{subequations} \label{eq:eta_r_recursion}
\begin{align}
\boldsymbol{\eta}_1^R (t+1)&= \boldsymbol{\eta}_1^R (t)+\gamma_{t+1} (
\boldsymbol{\mu}(t+1)- \boldsymbol{\eta}_1^R (t)) \label{eq:eta1_update} \\
\boldsymbol{\eta}_2^R (t+1)&= \boldsymbol{\eta}_2^R (t)+\gamma_{t+1} (\boldsymbol{\mu}(t+1)\odot|\boldsymbol{\phi}(t) |^2- \boldsymbol{\eta}_2^R (t)) \label{eq:eta2_update} \\
\boldsymbol{\eta}_3^R (t+1)&= \boldsymbol{\eta}_3^R (t)+\gamma_{t+1} (\boldsymbol{\mu}(t+1)\odot\boldsymbol{\phi}^*(t)- \boldsymbol{\eta}_3^R (t)) \label{eq:eta3_update} \\
\boldsymbol{\eta}_4^R (t+1)&= \boldsymbol{\eta}_4^R (t)+\gamma_{t+1} (\boldsymbol{\mu}(t+1)\odot\boldsymbol{\phi}(t)- \boldsymbol{\eta}_4^R (t)).\label{eq:eta4_update}
\end{align}
\end{subequations}
The maximization step in the \ac{REM} is similar to the maximization step in the batch \ac{EM}. The M-step $\boldsymbol{\theta}_{c,R}(t+1)=\argmax_{\boldsymbol{\theta_c}} \langle\boldsymbol{\eta}^R(\boldsymbol{\mu}(t+1), \boldsymbol{\phi}(t+1)),\boldsymbol{\epsilon}(\boldsymbol{\theta_c})\rangle$ yields the following parameter updates:
\begin{equation}
\psi_c(t+1) = \frac{1}{K} \sum_k \mu(t+1,k,c).
\label{eq:psi_rem_update} 
\end{equation}
The M-step and the E-step can be merged into one recursion:
\begin{subequations}
\label{eq:rem_updates}
\begin{align}
&\boldsymbol{\psi}_R(t) = \frac{1}{K} \sum_k \boldsymbol{\eta}_1^R(t) \label{eq:psi_update} \\
\intertext{and for the other parameters:}
&\tilde{\boldsymbol{\phi}}_R(\mathbf{p}(t))= \boldsymbol{\eta}_4^R(t) \oslash \boldsymbol{\eta}_1^R(t) \label{eq:phi_update} \\
&\begin{aligned} 
\boldsymbol{\sigma}_R^2(t) &= \frac{1}{M \cdot \sum_k \boldsymbol{\eta}_1^R(t,k)} \sum_{k,m} \Big( \boldsymbol{\eta}_1^R(t,k) \odot |\tilde{\boldsymbol{\phi}}_m^k(\mathbf{p}(t))|^2 \\
&\quad + \boldsymbol{\eta}_2^R(t,k,m) - \boldsymbol{\eta}_3^R(t,k,m) \cdot\tilde{\boldsymbol{\phi}}_m^k(\mathbf{p}(t)) \\
&\quad - \boldsymbol{\eta}_4^R(t,k,m) \cdot\tilde{\boldsymbol{\phi^*}}_m^k(\mathbf{p}(t)) \Big),
\end{aligned} \label{eq:sigma_update}
\end{align}
\end{subequations}
with $\oslash$ is an element-wise division operation. 
To infer the speaker's physical 2D position in Cartesian coordinates, denoted as $\hat{\mathbf{p}}(t)$, from the recursively updated parameters, a spatial grid search is performed at each time step. The estimated position is obtained by minimizing the squared error between the estimated \ac{PRP} and the theoretical PRP:
\begin{equation}
    \hat{\mathbf{p}}(t) 
    = 
    \arg\min_{\mathbf{p}}
    \sum_{k,m} 
    \left| 
        \tilde{\phi}_{R,m}^k(\mathbf{p}(t)) 
        - 
        \tilde{\phi}_m^k(\mathbf{p})
    \right|^2,
\end{equation}
where $\tilde{\phi}_m^k(\mathbf{p})$ is the theoretical value for a candidate location $\mathbf{p}$ as defined in~\eqref{eq:PRP}.

The performance of the \ac{CREM} algorithm is highly dependent on the choice of the step size $\gamma_t$. Classic stochastic approximation results guarantee convergence to a (static) local optimum given $\sum \gamma_t = \infty$ and $\sum \gamma_t^2 < \infty$. A common schedule is $\gamma_t = (t+2)^{-\alpha}$, where $0.5 < \alpha \le 1$ balances the trade-off between convergence speed (smaller $\alpha$) and estimation stability (larger $\alpha$). To further enhance stability, we can adopt a mini-batch approach of size $B$, which interpolates between pure online ($B=1$) and batch ($B=T$) estimation \cite{Liang2009OnlineEM}. However, since fixed schedules struggle with dynamic trajectories, we propose an unfolded architecture that learns an adaptive, data-driven step size, as described in the next section.

\begin{figure}[t]
    \centering
    \includegraphics[width=0.95\columnwidth]{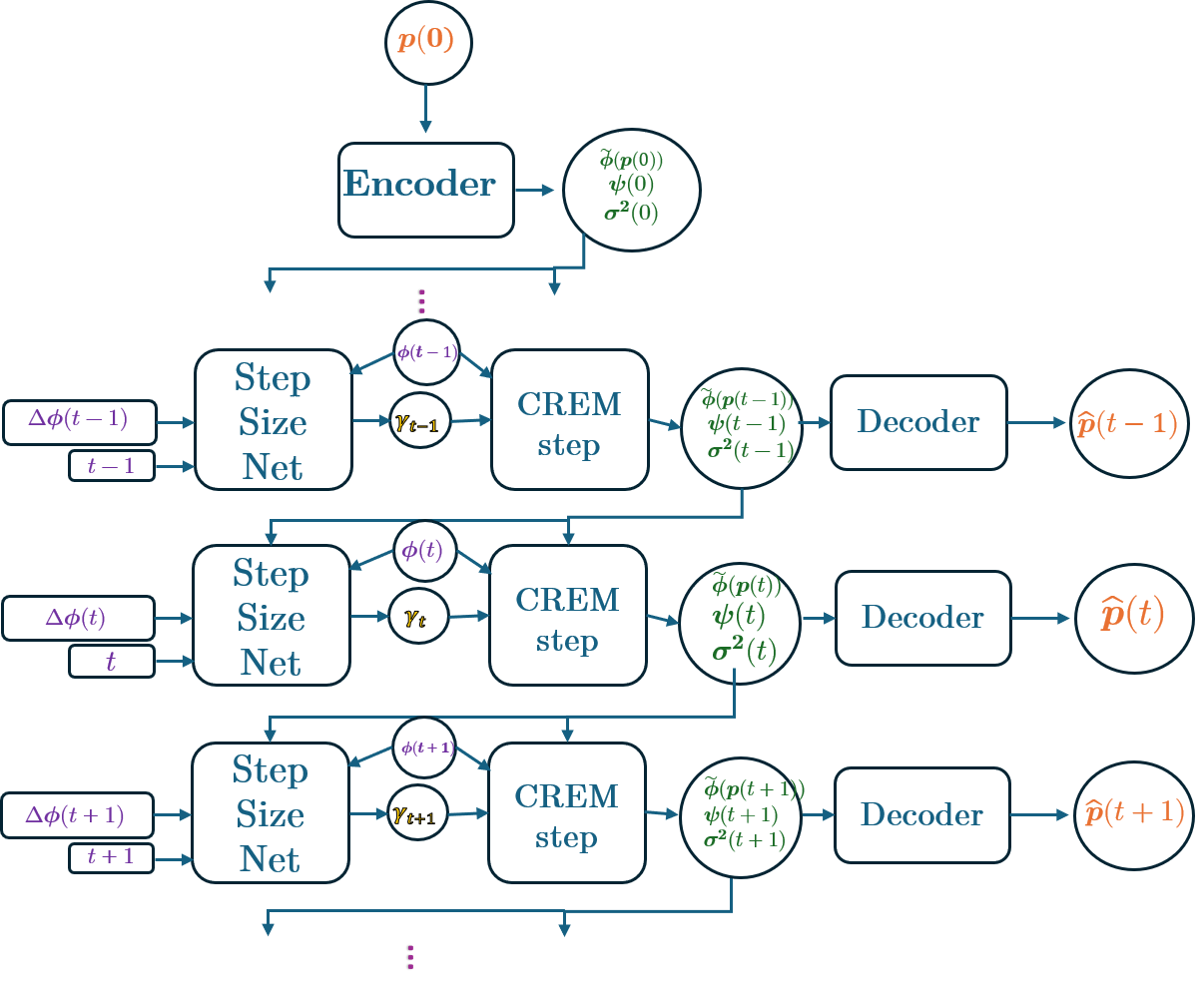}
    \centerline{\footnotesize (a) Full Unfolded CREM Architecture}
    
    \vspace{3pt} 
    
    \includegraphics[width=0.75\columnwidth]{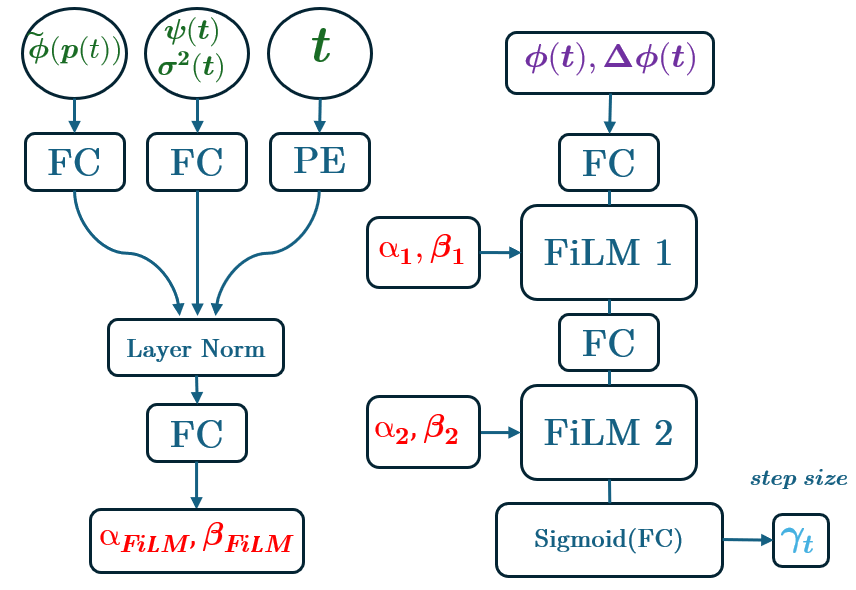}
    \centerline{\footnotesize (b) Step-Size Network}
    \addtolength{\belowcaptionskip}{-9pt}
\addtolength{\abovecaptionskip}{-6pt}
    \caption{Overview of the proposed tracking framework: (a) shows the unrolled EM layers within the encoder-decoder structure, and (b) details the FiLM-based adaptive step-size generation.}
    
    \label{fig:total_network}
\end{figure}

\vspace{2pt}\noindent\textbf{Unfolding CREM Neural Network:}
Building upon the interpretable Batch-EM unfolding framework presented in our previous work \cite{RinaVeler2026UnfoldedEM}, we propose a deep unfolded architecture for the \ac{CREM} algorithm. The network embeds the recursive EM steps within an encoder-decoder structure and introduces a data-driven mechanism to learn the optimal time-varying step size $\gamma_t$ (\autoref{fig:total_network}).

\noindent\textit{a) Encoder-decoder framework:} The core of the architecture consists of an encoder and decoder. The encoder, implemented as a \ac{FC} network, maps candidate spatial positions to an initial \ac{PRP} vector. The decoder performs the inverse mapping, translating the refined \ac{PRP} estimates back to spatial coordinates $\hat{\mathbf{p}}(t)$. By using the decoder at each time step, the network provides online tracking of the speaker's trajectory. To enhance stability, each layer processes mini-batches of size $B$ by interpolating between the online and batch regimes.

\noindent\textit{b) Dynamic step-size learning via \ac{FiLM} and \ac{PE}:} The primary innovation in this architecture is the Step-Size Network, which replaces the heuristic schedules of $\gamma_t$ with a learned, adaptive function. Since the optimal update step depends on both the current convergence state and the temporal context, we employ \ac{FiLM} \cite{perez2018film} and \ac{PE} \cite{vaswani2017attention}.
To provide the network with information about the current iteration, we use a sinusoidal positional encoding, $\textrm{PE}(t)$.The step-size network receives a conditioning vector $\mathbf{v}_t = [\textrm{PE}(t), \boldsymbol{\psi}(t-1), \boldsymbol{\sigma}^2(t-1), \boldsymbol{\phi}(t-1)]^\top$, which is processed through a Layer-Normalized \ac{MLP} to generate the \ac{FiLM} parameters, $\alpha$ (scaling) and $\beta$ (shifting).
The intermediate features are extracted from the concatenation of the current observation $\boldsymbol{\phi}(t)$ and the temporal difference between successive observations $\Delta \boldsymbol{\phi}(t) = \boldsymbol{\phi}(t) - \boldsymbol{\phi}(t-1)$. Then, the \ac{FiLM} parameters modulate the intermediate features to produce the scalar step size $\gamma_t$ used in \eqref{eq:eta_recursion}. The architecture of the step-size network is shown in \autoref{fig:total_network}.

\noindent\textit{c) Weighted loss:} The network is trained using a time-weighted composite loss function. To account for the initial convergence period of the \ac{REM}, we introduce a warmup weight $w(t)$ that scales the loss at each time step. The spatial accuracy is governed by a logarithmic distance error, which provides a sub-linear penalty to stabilize gradients against large initial offsets:
\begin{equation}
\begin{split}
\mathcal{L} = &\frac{1}{T} \sum_{t=0}^{T-1} w(t) \Big[ (1-\lambda) \log(1 + |\hat{\mathbf{p}}(t) - \mathbf{p}_{\text{true}}(t)|^2)  \\
&+ \lambda \cdot (1 - \text{CosSim}(\tilde{\boldsymbol{\phi}}(\mathbf{p}(t)), \tilde{\boldsymbol{\phi}}(\mathbf{p}_{\text{true}}(t)))\Big]
\end{split}
\label{eq:total_weighted_loss}
\end{equation}
where $\text{CosSim}(\cdot, \cdot)$ denotes the cosine similarity operator.
%
\section{Experimental Study}
\label{Experimental study}

\vspace{2pt}\noindent\textbf{Data Generation and Setup:}
\label{Signals generation}
We evaluated the proposed unfolded network using a synthetic dataset derived from the \ac{WSJ} corpus \cite{Paul1992Design}. The simulations were implemented in Python using the image-method-based signal generator \cite{Habets2006RoomImpulse}. We simulated a single moving speaker in rectangular rooms with random dimensions (height: 2.2–2.6 m; width/length: 5–7 m) with eight microphone pairs (intra-pair distance: 0.2 m).
The speaker moves at a constant velocity of $0.5$~m/s, following either a linear trajectory or a circular path with a radius ranging from $1$ to $3$~m. 
Acoustic conditions include either anechoic or reverberant environments with $\textrm{RT}_{60} =0.3$~s and additive white noise at $\text{SNR}=30$~dB. The dataset consists of $8{,}000$ training and $2{,}000$ validation samples.

\vspace{2pt}\noindent\textbf{Signal Processing and Data Flow:}
The input signals, sampled at $16$~kHz, feature durations of $5$~seconds in anechoic scenarios and $9$~seconds in reverberant conditions. Transformed into the \ac{TF} domain using an \ac{STFT} (window size of $1024$, hop size of $512$ samples), these configurations yield $157$ and $282$ time frames, respectively. To optimize efficiency, only frequency bins within the $500-1500$~Hz range are considered. Tracking is performed with a mini-batch size of $B=10$ consecutive time frames at each step, employing a sliding window with a step size of $3$ frames between consecutive \ac{CREM} updates to maintain temporal continuity.

\vspace{2pt}\noindent\textbf{Encoder and Decoder Architectures:}
The \ac{FC} encoder serves as the initialization block, mapping random room coordinates to the initial \ac{PRP} vector through $7$ layers with increasing widths: $[6, 48, 192, 384, 768, 1536, 2080]$. Each layer is followed by LayerNorm and \ac{ReLU} activation, while the final output uses a \textit{tanh} activation to ensure the stability of the \ac{PRP} values. 
The decoder, which maps the refined \ac{PRP} back to spatial coordinates, comprises $6$ layers with decreasing dimensions: $[1040, 512, 256, 128, 64, 2]$. All layers utilize \ac{ReLU} activation, except for the final output layer, which provides the continuous $2$D coordinates.

\vspace{2pt}\noindent\textbf{Step-Size Network with \ac{FiLM}:}
The adaptive step size $\gamma_t$ is estimated using a multi-branch conditioning network. 
(i) \textit{Parameter Branch:} $\boldsymbol{\psi}(t)$ and $\boldsymbol{\sigma}^2(t)$ are mapped to a $16$-dimensional feature; 
(ii) \textit{Manifold Branch:} The estimated $\tilde{\boldsymbol{\phi}}(\mathbf{p}(t))$ passes through $4$ layers ($[512, 256, 128, 64]$) with \ac{ReLU}; 
(iii) \textit{Temporal Branch:} The time index $t$ is encoded via a $32$-dimensional \ac{PE}.
These three branches are concatenated and passed through LayerNorm to generate the \ac{FiLM} parameters ($\alpha, \beta$). Two separate generators are used: a $3$-layer \ac{FC} ($[256, 512, 1024]$) for a $512$-dim modulation, and a $2$-layer \ac{FC} ($[164, 256]$) for a $128$-dimensional modulation.

The main step-size estimation path concatenates the current observation $\boldsymbol{\phi}(t)$ and the temporal difference $\Delta \boldsymbol{\phi}(t)$. This input is processed through a $512$-dimensional \ac{FC} layer, followed by the first \ac{FiLM} block, two \ac{FC} layers ($[256, 128]$), the second \ac{FiLM} block, and finally $3$ \ac{FC} layers ($[64, 32, 1]$). A \textit{sigmoid} activation is applied to the final scalar output to bound the learned step size $\gamma_t \in [0, 1]$.

\vspace{2pt}\noindent\textbf{Training and Stability:}
Training was performed using the composite loss in \eqref{eq:total_weighted_loss} with $\lambda=0.5$. To handle the initial convergence period, we employed a linear warmup schedule for the loss weights $w(t)$, increasing from $0.1$ to $1.0$ over the first $20$ time steps and remaining constant thereafter. To ensure numerical stability during the unfolding of recursive updates, we applied several regularization techniques:
(i) gradient clipping with a threshold of $1.5$; 
(ii) weight decay of $10^{-4}$; 
(iii) a clamp value of $50$ on the estimated variances $\sigma^2$; 
and (iv) a small epsilon $\epsilon = 10^{-8}$ added to all denominators in the \ac{CREM} update rules. The network was optimized using Adam with an initial learning rate of $10^{-3}$.

\vspace{2pt}\noindent\textbf{Results and Discussion:}
\label{Results}
We evaluated the proposed approach against the \ac{CREM} baseline using $100$ unseen speech signals. To demonstrate the impact of our learned model, the baseline performance was tested across multiple fixed step sizes, specifically evaluating $\gamma_t \in \{0.25, 0.5, 0.75\}\ \forall \,t$, coupled with a spatial grid search. These evaluation scenarios incorporated diverse room sizes and varying trajectory locations to assess localization accuracy, tracking adaptability, and robustness across different acoustic conditions.
\autoref{tab:results} summarizes the results. We report the warmup-weighted average \ac{RMSE} between the estimated and ground-truth speaker positions, along with the percentage of test trajectories for which the time-averaged positional error exceeds $0.5$~m.
\begin{table}[htbp]
\centering
\setlength{\belowcaptionskip}{-8pt}
\setlength{\abovecaptionskip}{2pt}
\caption{Average Localization Performance: Unfolded Network vs. \ac{CREM} Baseline.}
\label{tab:results}
\resizebox{1.0\columnwidth}{!}{
\begin{tabular}{lcccc}
\toprule
\multirow{2}{*}{\textbf{Rever. Level}} & \multicolumn{2}{c}{\textbf{Unfolded \ac{CREM} Network}} & \multicolumn{2}{c}{\textbf{\ac{CREM} Baseline ($\gamma = [0.25, 0.5, 0.75]$)}} \\
\cmidrule(lr){2-3} \cmidrule(lr){4-5}
& \textbf{\ac{RMSE} (m)} & \textbf{\textgreater{}0.5 m (\%)} & \textbf{\ac{RMSE} (m)} & \textbf{\textgreater{}0.5 m (\%)} \\
\midrule
$T_{60} = 0$~s   & \textbf{0.25} & 3  & [0.28, 0.67, 1.44] & [\textbf{1}, 29, 61] \\
$T_{60} = 0.3$~s & \textbf{0.55} & \textbf{56} & [0.74, 0.81, 0.86] & [87, 93, 97] \\
\bottomrule
\end{tabular}}
\vspace{-4pt}
\end{table}
Under anechoic conditions, the unfolded network outperforms the baselines with larger fixed step sizes while achieving performance comparable to the $\gamma=0.25$ baseline. This is consistent with the learned step size dynamically converging to a similar low value. In the more challenging reverberant scenario ($T_{60}=0.3$~s), the proposed method continues to outperform all fixed-step baselines, although its performance degrades relative to the anechoic case. The advantage over the $\gamma=0.5$ baseline is further illustrated by the error distribution in Fig.~\ref{fig:error_hist}, where the unfolded network (blue) exhibits a substantially higher concentration of trajectories with low \ac{RMSE} than the baseline (orange).
\begin{figure}[t]
    \centering
    \includegraphics[width=0.9\columnwidth]{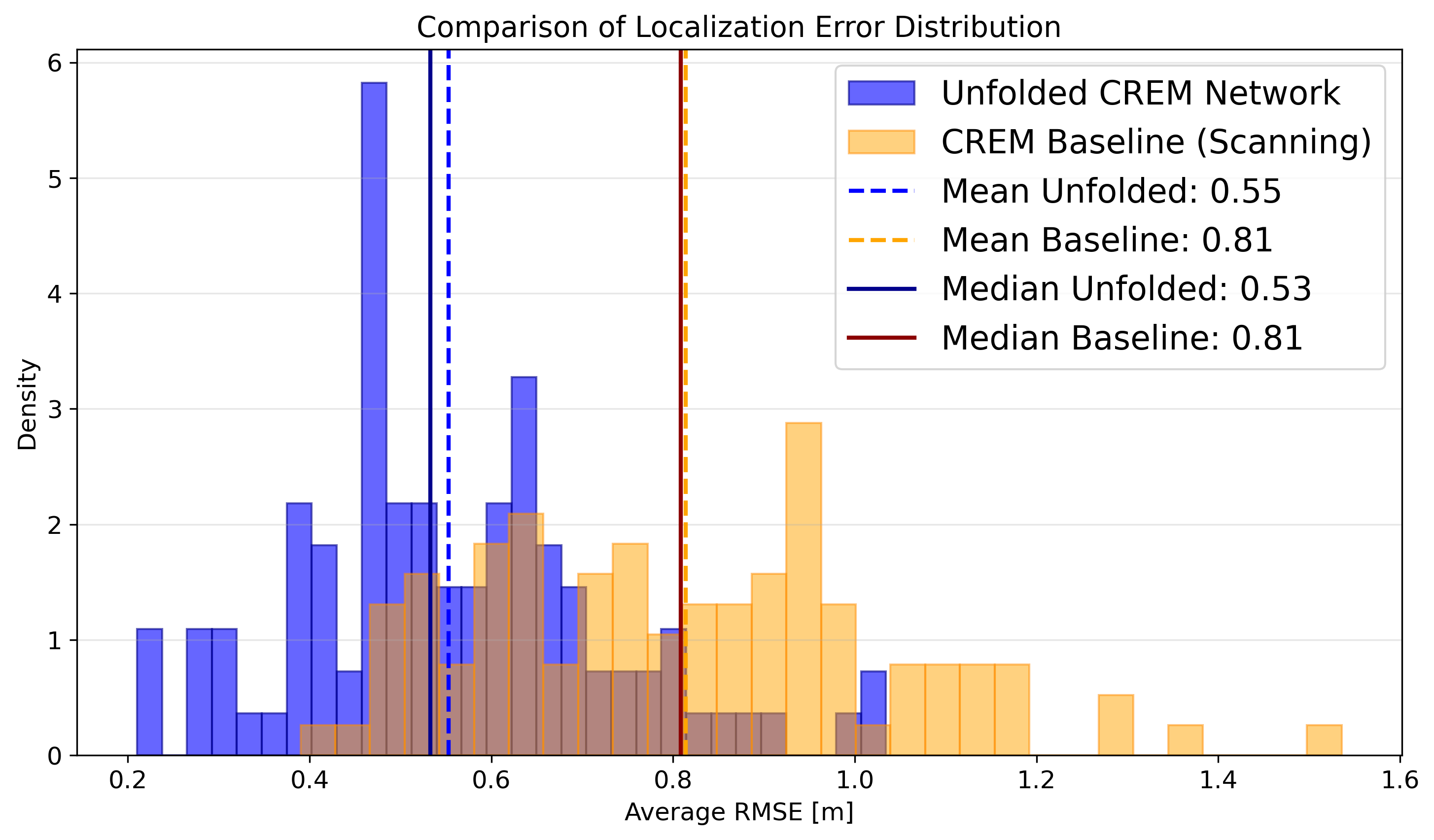} 
        \addtolength{\belowcaptionskip}{-9pt}
\addtolength{\abovecaptionskip}{-4pt}
    \caption{Comparison of localization \ac{RMSE} distribution between the unfolded network and the \ac{CREM} baseline ($\gamma_t = 0.5$). The proposed method's error distribution is more concentrated at lower values; median (solid) and mean (dashed) lines are provided to illustrate the performance gap.}
    \label{fig:error_hist}
\end{figure}
\begin{figure}[h!] 
\vspace{-14pt}
    \centering
    \begin{subfigure}{0.48\columnwidth} 
    \centering\includegraphics[width=0.95\textwidth,  keepaspectratio]{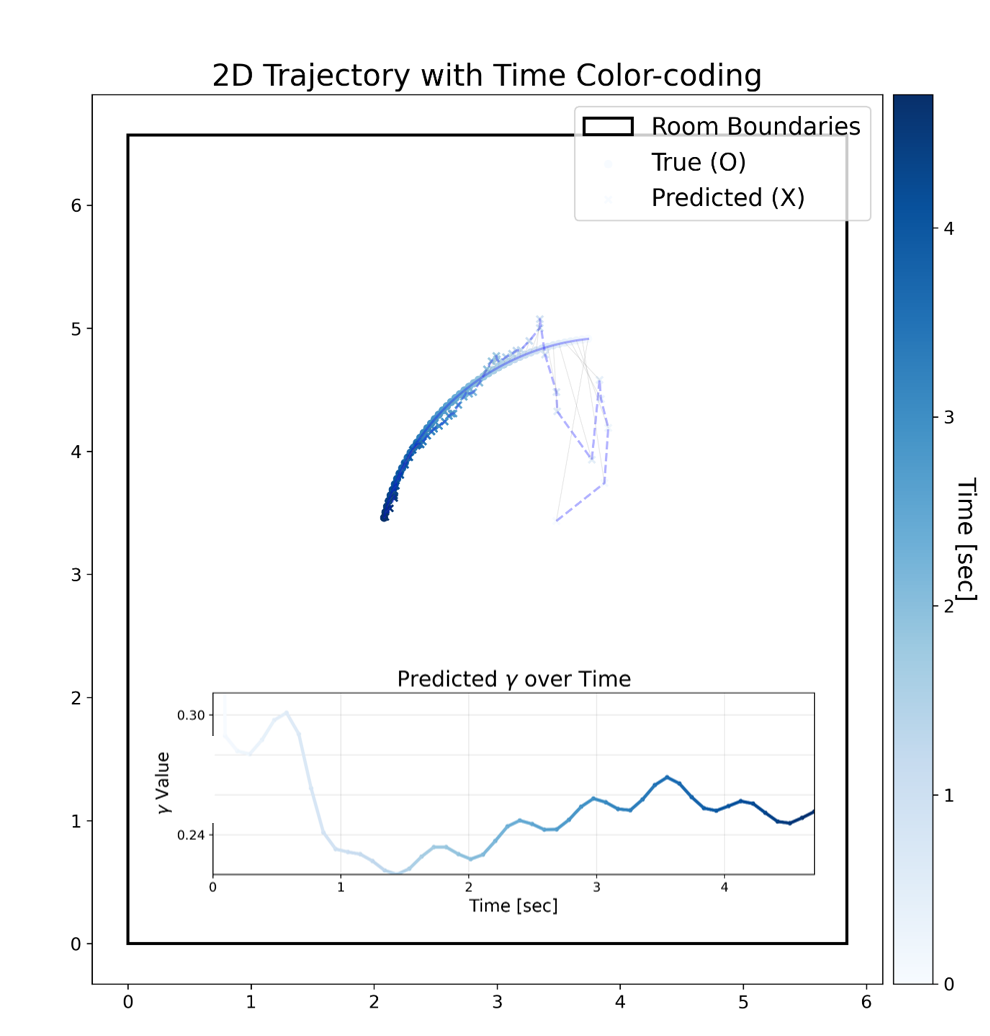}
        \caption{Speaker tracking in an anechoic room ($\textrm{RT}_{60} = 0$~s). Average \ac{RMSE} $=0.09$~m}
        \label{fig:tracking_anechoic}
    \end{subfigure}
    \hfill 
    \begin{subfigure}{0.48\columnwidth}
        \centering
        \includegraphics[width=0.99\textwidth, keepaspectratio]{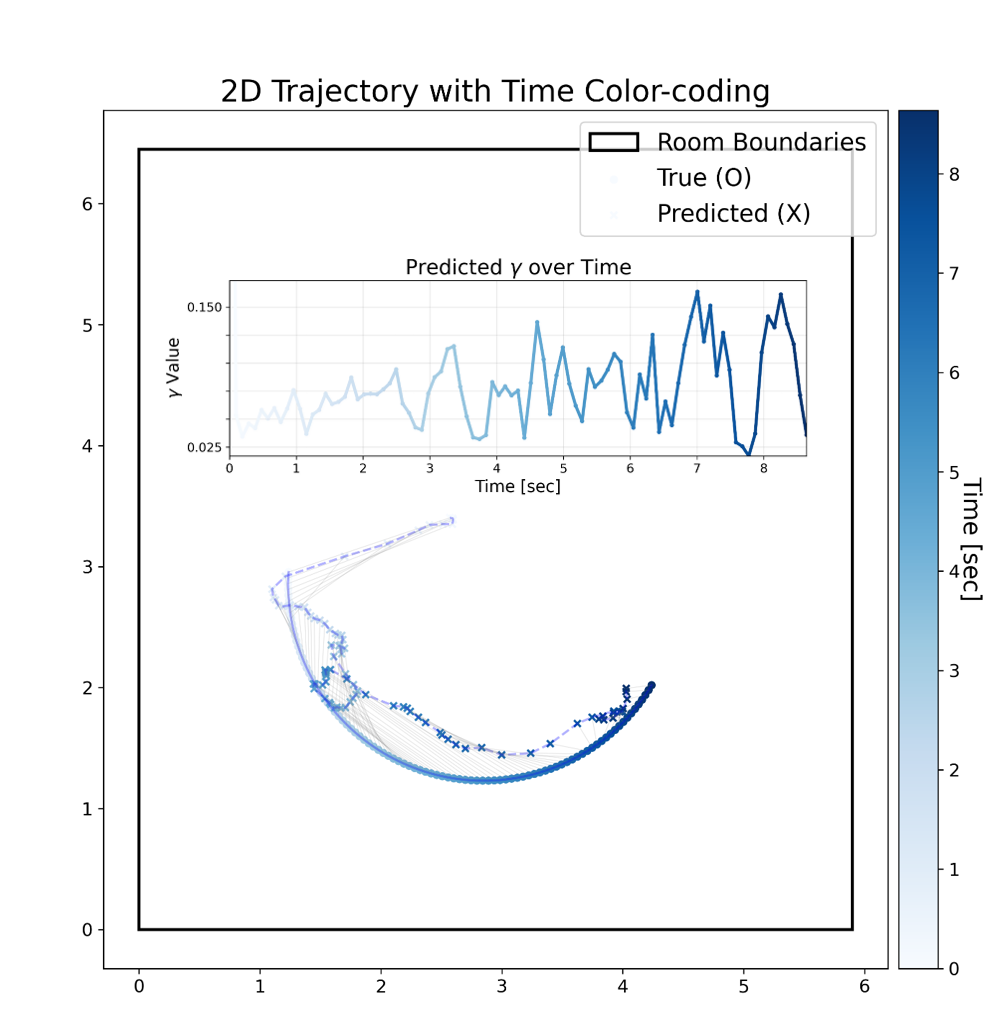}
        \caption{Speaker tracking in a reverberant room ($\textrm{RT}_{60} = 0.3$~s). Average \ac{RMSE} $=0.57$~m}
        \label{fig:tracking_reverberant}
    \end{subfigure}
    \addtolength{\belowcaptionskip}{-4pt}
\addtolength{\abovecaptionskip}{-3pt}
    \caption{Trajectory estimation examples from test data. True positions are marked by dots, and network estimates by `x` markers. The color gradient represents the progression of time and the accompanying plot displays the corresponding learned step-size values.}
    \label{fig:tracking_comparison}
\end{figure}
The network's tracking capabilities are further qualitatively illustrated in Fig.~\ref{fig:tracking_comparison}, which shows the estimated trajectories under both anechoic and reverberant conditions.

\section{Conclusion}
\label{Conclusion}

We presented an unfolded \ac{REM} network for speaker tracking that replaces fixed step-size schedules with a learned, adaptive update policy. Experimental results demonstrate that the proposed model outperforms the classical \ac{CREM} baseline, achieving lower \ac{RMSE} and improved stability in both anechoic and reverberant environments.
Analysis of the learned parameters shows that the step size $\gamma_t$ converges to a distinct range for each acoustic setting. In anechoic scenarios, $\gamma_t$ fluctuates around $0.25$ (peaking at $0.55$), whereas in reverberant conditions it decreases to $0.02$--$0.1$. The network learns to stabilize the step size, likely because the speaker velocity remains constant. Future work will focus on improving robustness under highly reverberant conditions and extending the model to more complex scenarios with variable speaker velocities and trajectories.
\balance
\bibliographystyle{IEEEbib}
\bibliography{refs25}

@string{icml = "Proc. ICML"}

@string{neurips = "Proc. NeurIPS"}

@article{Schwartz2014,
  author    = {Schwartz, Ofer and Gannot, Sharon},
  title     = {Speaker Tracking Using Recursive {EM} Algorithms},
  journal   = {IEEE/ACM Transactions on Audio, Speech, and Language Processing},
  volume    = {22},
  number    = {2},
  pages     = {392--402},
  year      = {2014},
  month     = feb,
  doi       = {10.1109/TASLP.2013.2292361},
  publisher = {IEEE}
}

@article{Chakrabarty2019MultiSpeaker,
  author    = {Chakrabarty, Soumitro and Habets, Emanuel A.},
  title     = {Multi-Speaker {DOA} Estimation Using Deep Convolutional Networks Trained With Noise Signals},
  journal   = {IEEE Journal of Selected Topics in Signal Processing},
  volume    = {13},
  number    = {1},
  pages     = {8--21},
  year      = {2019},
  publisher = {IEEE}
}

@inproceedings{Gregor2010Learning,
  author    = {Gregor, Karol and LeCun, Yann},
  title     = {Learning fast approximations of sparse coding},
  booktitle = {Proceedings of the International Conference on Machine Learning (ICML)},
  pages     = {399--406},
  year      = {2010},
  doi       = {10.5555/3104322.3104374},
  publisher = {ACM}
}

@article{Monga2021Algorithm,
  author    = {Monga, V. and Li, Y. and Eldar, Y. C.},
  title     = {Algorithm Unrolling: Interpretable, Efficient Deep Learning for Signal and Image Processing},
  journal   = {IEEE Signal Processing Magazine},
  volume    = {38},
  number    = {2},
  pages     = {18--44},
  year      = {2021},
  publisher = {IEEE},
  doi       = {10.1109/MSP.2020.3042412} 
}

@inproceedings{Greff2017NeuralEM,
  author    = {Greff, Klaus and van Steenkiste, Sjoerd and Schmidhuber, Juergen},
  title     = {Neural Expectation Maximization},
  booktitle = {Proceedings of the 31st Conference on Neural Information Processing Systems (NeurIPS)},
  pages     = {6694--6704},
  year      = {2017}
}

@inproceedings{Paul1992Design,
  author    = {Paul, Douglas B. and Baker, Janet M.},
  title     = {The design for the Wall Street Journal-based CSR corpus},
  booktitle = {Proceedings of the workshop on Speech and Natural Language},
  pages     = {357--362},
  year      = {1992},
  publisher = {Association for Computational Linguistics},
  organization = {ACL}
}

@article{Grumiaux2022Survey,
  author    = {Grumiaux, Pierre-Amaury and Kiti{\'c}, Sr{\dj}an and Girin, Laurent and Gu{\'e}rin, Alexandre},
  title     = {A survey of sound source localization with deep learning methods},
  journal   = {J. Acoust. Soc. Am.},
  volume    = {152},
  number    = {1},
  pages     = {107--151},
  year      = {2022},
  month     = jul
}

@inproceedings{Mandel2007EMAlgorithm,
  author    = {Mandel, Michael and Ellis, Daniel and Jebara, Tony},
  title     = {An {EM} algorithm for localizing multiple sound sources in reverberant environments},
  booktitle = {Advances in Neural Information Processing Systems},
  volume    = {19},
  pages     = {953},
  year      = {2007},
}

@article{Cappe2009OnlineEM,
  author    = {Capp{\'e}, Olivier and Moulines, Eric},
  title     = {On-line expectation-maximization algorithm for latent data models},
  journal   = {J. R. Statist. Soc.: Ser. B (Statist. Methodol.)},
  volume    = {71},
  number    = {3},
  pages     = {593--613},
  year      = {2009},
}

@article{Pu2023DeepEM,
  author    = {Pu, Yannan Pu and Sun, Jian and Tang, Niansheng and Xu, Zongben},
  title     = {Deep expectation-maximization network for unsupervised image segmentation and clustering},
  journal   = {Image and Vision Computing},
  volume    = {135},
  pages     = {104717},
  year      = {2023},
}

@inproceedings{Liang2009OnlineEM,
  author    = {Liang, Percy and Klein, Dan},
  title     = {Online {EM} for Unsupervised Models},
  booktitle = {Proceedings of Human Language Technologies: The 2009 Annual Conference of the North American Chapter of the Association for Computational Linguistics},
  address   = {Boulder, Colorado},
  pages     = {611--619},
  year      = {2009},
}

@techreport{Habets2006RoomImpulse,
  author  = {Habets, Emanuel A. P.},
  title   = {Room impulse response generator},
  institution = {Technische Universiteit Eindhoven},
  year    = {2006},
  note    = {[Online]. Available: https://www.audiolabs-erlangen.de/fau/professor/habets/software/rir-generator},
}

@article{Knapp1976GeneralizedCorrelation,
  author    = {Knapp, Charles H. and Carter, G. Clifford},
  title     = {The generalized correlation method for estimation of time delay},
  journal   = {IEEE Transactions on Acoustics Speech and Signal Processing},
  volume    = {24},
  number    = {4},
  pages     = {320--327},
  year      = {1976},
}

@article{Schmidt1986Multiple,
  author    = {Schmidt, Ralph},
  title     = {Multiple emitter location and signal parameter estimation},
  journal   = {IEEE Transactions on Antennas and Propagation},
  volume    = {34},
  number    = {3},
  pages     = {276--280},
  year      = {1986},
}

@article{RinaVeler2026UnfoldedEM,
  title={Speakers Localization Using Batch {EM} In Unfolding Neural Network},
  author={Veler Rina and Gannot Sharon},
  journal={arXiv preprint arXiv:2603.16278},
  year={2026},
  url={https://arxiv.org/abs/2603.16278}
}

@inproceedings{perez2018film,
  title={{FiLM}: Visual Reasoning with a General Conditioning Layer},
  author={Perez, Ethan and Strub, Florian and De Vries, Harm and Dumoulin, Vincent and Courville, Aaron},
  booktitle={Proceedings of the AAAI Conference on Artificial Intelligence},
  volume={32},
  number={1},
  year={2018},
  month={April}
}

@inproceedings{vaswani2017attention,
  title={{Attention} Is All You Need},
  author={Vaswani, Ashish and Shazeer, Noam and Parmar, Niki and Uszkoreit, Jakob and Jones, Llion and Gomez, Aidan N and Kaiser, {\L}ukasz and Polosukhin, Illia},
  booktitle={Advances in Neural Information Processing Systems},
  volume={30},
  pages={5998--6008},
  year={2017}
}
\end{document}